\documentclass[prc,twocolumn,english,showpacs]{revtex4}
\usepackage[T1]{fontenc}
\usepackage[latin1]{inputenc}
\usepackage{babel}
\usepackage{graphics}
\usepackage{comment}

\makeatletter

\usepackage[T1]{fontenc}
\usepackage[latin1]{inputenc}
\usepackage{babel}
\usepackage{graphics}



\makeatother
\begin{document}

\title{Transfer to the continuum and Breakup reactions}
\author{A.M. Moro}
\email{moro@us.es}
\affiliation{Departamento de F\' {\i}sica, 
Instituto Superior T\'ecnico, Tagus Park, 2780--990 Oeiras, Portugal}
\affiliation{Departamento
de FAMN, Universidad de Sevilla,
Apdo. 1065, E-41080 Sevilla, Spain}

\author{F.M. Nunes}
\email{nunes@nscl.msu.edu}
\affiliation{NSCL and Department of Physics and Astronomy, 
Michigan State University, U.S.A.}

\begin{abstract}
Reaction theory is an essential ingredient when performing
studies of nuclei far from stability. 
One approach for the calculation of breakup reactions of exotic
nuclei into two fragments is to consider inelastic excitations into
the single particle continuum of the projectile. Alternatively
one can also consider the transfer to the continuum of a system composed
of the light fragment and the target. In this work we 
make a comparative study of the two
approaches, underline the different inputs, and identify the
advantages and disadvantages of each approach. Our test
cases consist of the breakup of $^{11}$Be on a proton target at
intermediate energies, and the breakup of $^8$B on $^{58}$Ni
at energies around the Coulomb barrier. We find that,
in practice the results obtained in both schemes
are in semiquantitative agreement.
We suggest a simple condition that can select between the
two approaches.
\end{abstract}


\date{\today}
\maketitle

\section{Introduction}
A large fraction of present nuclear physics encompasses the study 
of nuclear structure far from stability using reaction measurements. 
Extraction of fundamental structure therefore requires adequate 
reaction theories. As one of the most important tools, breakup offers a 
unique opportunity to benchmark reaction theories.
From the early days it became clear that the standard models to breakup
of stable nuclei needed revision \cite{jpg-rev}. Many of
the lessons learnt from the deuteron breakup have since become
a source of inspiration for the rare isotope reaction community \cite{Tos98a}. 
Different groups designed various reaction models, tailored to specific
systems and using particular approximations. Albeit the variety, the 
state of the art of the existing reaction-model panorama has become 
increasingly unsatisfying:
at present we already have a handful of models that produce results
for a specific case but we are missing a general effort of a
consistent comparison between the various approaches. In the few cases
where two different models are applied to the same problem, there is
often a disparity in the predictions 
\cite{Win01,Tim99,Mort02,Sch03,Mar02,Nun99}. 
It is timely to make the necessary links between the available models.
Some studies, comparing approximations
such as eikonal, adiabatic, local momentum, have been recently performed
\cite{Bon01,Shy01,Zad04}. Here, we work within a framework
where no such approximations are present.

Roughly speaking, current breakup reaction theories can be divided into
two main categories. On one side, some methods model the breakup
process as an excitation of the projectile to the continuum spectrum of the 
projectile. This is the case of the CDCC \cite{Yah86,Aus87} approach. On
the other side, some other methods, such as the semiclassical transfer to the 
continuum developed by Brink and Bonaccorso \cite{Bon88,Bon91a,Bon91b,Bon92,Bon01,Gar05} 
or the post-form DWBA 
approach \cite{Baur75,Ban98,Chat00,Chat02}, treat the
breakup process as the transfer of 
one of the fragments to the unbound states of the target. Intuitively,
it is obvious that both continua do correspond to the same three-body
continuum, expressed in different coordinates systems. However, it is not
clear to what extent this equivalence is fulfilled in a practical calculation. In this
work we try to shed some light on this problem by applying both approaches
to the same reaction and using, whenever possible, the same physical ingredients. 

Although many kinds of observables could be calculated in both methods, we make
special emphasis on core energy and angular distributions, since these observables
are particularly important in currently measured reactions with radioactive beams.


The paper is organized as follows. In section II we briefly review 
the three-body breakup and transfer to the continuum approaches,
in the form used in this work. In Sec.\ III, we
apply these formalisms to the reactions  $p$+$^{11}$Be and $^8$B+$^{58}$Ni. A
discussion of these results is presented in Sec.\ IV and conclusions are 
drawn in Sec.\ V.


\begin{figure}[b]
\resizebox*{0.3\textwidth}{!}{\includegraphics{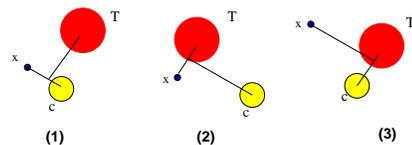}}
\vspace{-0.4cm}
\caption{\label{jacobi} (Color online) The three Faddeev components 
for the problem
of a two-body projectile ($c+x$) impinging on a target $T$.}
\end{figure}

\section{Three-body reaction models}
When considering reactions with light radioactive beams, it is
customary to model the incoming projectile as a two-body system
(in fact sometimes the projectile has a clear three-body structure
and models that handle three-body projectiles are underway).
In principle, the solution to the reaction problem can be obtained exactly
from solving the Faddeev equations with the appropriate boundary conditions.
In the Faddeev formalism \cite{Fad60,Joa87}, the three-body 
wavefunction is written as
a sum of three Jacobi components represented in 
Fig.~\ref{jacobi}. Each component is defined by the intercluster coordinate
between two of the subsystems (\{${\bf r}_i;i=1,2,3$\}) and 
the relative coordinate
of this pair to the third cluster (\{${\bf R}_i; i=1,2,3$\}).
Asymptotically, the Faddeev component $i$
contains contribution from the bound states associated to the
pair with relative coordinate ${\bf r}_i$, 
plus a contribution coming from three-body breakup. Therefore, while 
rearrangement channels are confined to specific Faddeev components,
breakup is distributed among the three components. Consequently,
extracting
the breakup observables requires complicated transformations among Jacobi 
coordinates. Besides, solving the Faddeev coupled equations is a very 
difficult task, specially
for three charged particles where Coulomb plays an important role. Inclusion
of absorption in the intercluster potentials, which is required when the 
subsystems have internal degrees of freedom is also an open problem, although
some promising work is already in progress \cite{akram}.

It is well known that basis states belonging to different Jacobi sets
are not mutually orthogonal. Furthermore, for each Jacobi set, a 
complete basis of the three-body bound and unbound spectrum
can be constructed.   Then, it could be possible, in principle, to describe the reaction
observables of a thee-body scattering problem using uniquely states from one of the 
Jacobi components.  This is in fact the procedure followed by the Continuum Discretized 
Coupled Channels (CDCC) formalism. This method has been applied for more than two
decades to the scattering of weakly bound (two-body) 
projectiles by light and heavy targets.
%
For the scattering of a composite 
projectile $A$(=$c$+$x$) by a target $T$, the CDCC method defines the 
model three-body Hamiltonian:
\begin{eqnarray}
H &=& K_{rel} + H_{int} + U_x + U_c, \nonumber \\
H_{int} &=& K_{int} + V_{xc}
\label{H-cdcc}
\end{eqnarray}
where $K_{rel}$  is the kinetic energy for the projectile-target 
relative motion, $K_{int}$ is the internal kinetic energy of the 
projectile, $U_x$ and $U_c$ are the $x-T$ and $c-T$ interactions 
and $V_{xc}$ is the
$x-c$ binding potential.  As to the internal degrees of freedom of the
target, in the standard CDCC method, only the target ground
state is considered explicitly. 
Therefore, the fragments are not allowed 
to engage in arbitrary processes with the target.
For example, processes in 
which one of the dissociated fragments is absorbed by the target,
or in which the target internal degrees of freedom are excited, 
are excluded from the model space. Also, rearrangement channels
corresponding to cluster-target bound states are by construction excluded 
from the CDCC model space and hence, those observables associated with these
two-body channels can not be obtained from the asymptotics of the 
CDCC three-body wavefunction. The model space spanned by CDCC 
allows only the calculation of {\it elastic breakup} and leaves out 
those processes related to {\it inelastic breakup}. 

In order to take into account the effect of the excluded channels,
the interactions $U_x$ and $U_c$ are usually 
taken as phenomenological optical potentials obtained, for example,
from the fit of the elastic data 
at the same energy per nucleon. 
By contrast,
the interaction $V_{xc}$ is taken to be real, and 
chosen to reproduce known bound and/or excited
states separation energies, or resonance energies.


The full three-body space is truncated   by setting a maximum 
excitation energy for the projectile. Moreover, the $c-x$ relative angular
momentum is also restricted by considering only a limited number of partial 
waves. In order to deal with a finite set of coupled
equations, a discretization of the continuum states into energy intervals
(bins) is
also performed. This procedure should be regarded as a practical method
of making the problem numerically solvable, rather than an additional 
approximation. In fact, it has been shown that
the calculated observables are essentially independent of the method 
of discretization \cite{Piy99}.

Within this restricted model space, the three-body wavefunction is 
expanded in eigenstates of the internal 
Hamiltonian $H_{int}$ as
\begin{eqnarray}
\Psi_{\bf K}^{(1)}({\bf r}_1,{\bf R}_1) = \sum_{\alpha=0}^{N_1}
\phi_{\alpha}({\bf r}_1) \chi_{\alpha}({\bf R}_1),
\label{cdccwf}
\end{eqnarray}
where $N_1$ is the number of states considered, 
$\alpha$ represents all angular momentum quantum numbers as well
as excitation energies of the projectile, 
$\phi_{\alpha}({\bf r})$ are the eigenstates of the two-body Hamiltonian
$H_{int}$ and $\chi_{\alpha}({\bf R})$ describes the relative
motion between the projectile $A=c+x$ and the target $T$. This  expansion of the three-body
 wavefunction is inserted into the Shr\"odinger equation
that, when projected into the considered internal states, provides  a 
set of coupled equations.

\begin{figure}[t]
\resizebox*{0.23\textheight}{!}{\includegraphics{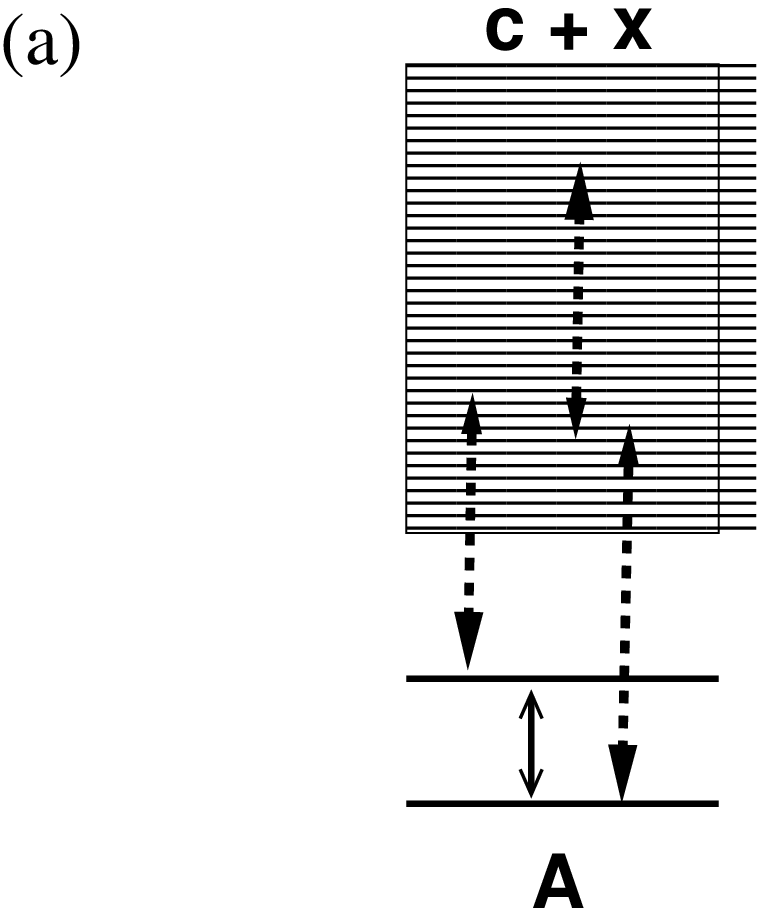}}
\\
\vspace{0.5cm}
\resizebox*{0.2\textheight}{!}{\includegraphics{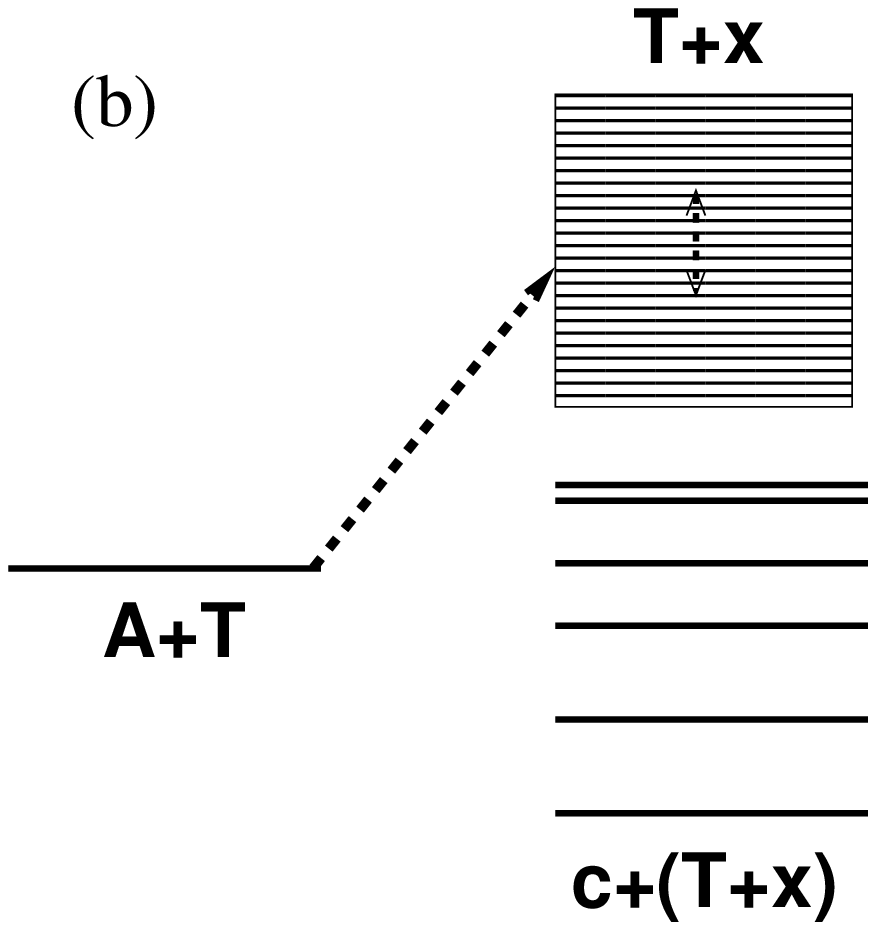}}
\vspace{-0.4cm}
\caption{\label{coupbu}: 
(a) Breakup couplings for a two-body projectile 
($c+x$) impinging on a target $T$ and (b) corresponding
transfer to the continuum couplings.}
\end{figure}

Within the CDCC scheme, the breakup 
process is treated as inelastic
excitations of the projectile $A$ into the continuum $c+x$ due to
the interactions with the target $T$ \cite{Yah82,Yah86,Aus87}. A pictorial 
representation of these couplings is given in Fig.~\ref{coupbu}(a). The
couplings responsible for this excitation, as well as the diagonal potentials,
are obtained by folding the phenomenological interactions $U_x$ and $U_c$ with
the internal wavefunctions, i.e.
\begin{eqnarray} \label{eqcoup1}
U_{\alpha;\alpha'}({\bf R}_1) & = & 
\langle \phi_\alpha|U_{x} + U_{c}|\phi_{\alpha'}\rangle\;.
\end{eqnarray}

Applications to the breakup of $^8$B at low and intermediate energy 
regimes have been very successful in describing the data 
\cite{Nun99,Tos01,Mort02}. 

Unlike the Faddeev method, the CDCC approach uses only one of the three
possible sets of Jacobi coordinates. As noted above, rearrangement channels
corresponding to cluster-target bound states are not part 
of the CDCC model space and, therefore, the 
CDCC three-body wavefunction is not adequate to predict observables 
associated with these two-body channels. 
On the contrary, it has been argued 
that, provided that the model space is sufficiently large \cite{Aus89,Aus96},
the total three-body breakup is contained in the CDCC wavefunction and, hence,
can be extracted from its asymptotics.

Although considerably simpler than its Faddeev counterpart,  
solving the CDCC problem is also a 
complicated task. In some cases, particularly with heavy targets, long range
interactions usually lead to convergence problems.
An additional difficulty is that, in many breakup experiments, 
scattering observables (differential energy cross sections, angular
distributions, etc) are obtained with respect to one of the projectile
fragments (this is indeed always the case of inclusive reactions). Given 
the choice of coordinates, the CDCC observables are
more naturally expressed in terms of the projectile center of mass, and
its internal excitation energy. Converting
to one of the fragment's coordinates requires a complicated kinematic
transformation \cite{Tos01}.
Furthermore, due to the restricted
model space, the CDCC description is not expected to be good  in the 
region  where channels outside the model space play an important role. Discrepancies
in large angle scattering data, observed in early applications of the CDCC
method to deuteron and $^{3}$He breakup \cite{Ise86}, have been attributed to this
fact. 

A way to circumvent the two latter criticisms is to use the T-matrix formalism
in post-form and approximate the incoming exact three-body wavefunction appearing
in the exact scattering amplitude:
\begin{equation} 
T_{post}=\langle \chi^{(-)}_{cB} \phi^{(-)}_{xT} |V_{xc}+U_{c}-U_f|\Psi_{i}^{(+)}\rangle ,
\label{Tpost1}
\end{equation}
by the CDCC wavefunction, i.e., $\Psi _{i}^{(+)}\approx  \Psi^{CDCC}$.  In this equation, $B=x+T$, 
$\chi^{(-)}_{cB}$ is the distorted wave generated by the (arbitrary) distorting potential
$U_f(\mathbf{R'})$ (where $\mathbf{R'}$ is the $c-B$ relative coordinate), and $\phi^{(-)}_{xT}$ 
represents a scattering state for the $x+T$ system.
By making use of the
Gell-Mann--Goldberger two-potential formula, the transition 
amplitude (\ref{Tpost1}) can be rewritten as:
\begin{equation} 
T_{post}=\langle\chi^{(-)}_{cT}\phi^{(-)}_{xT}  |V_{xc}| \Psi _{i}^{(+)}\rangle ,
\label{Tpost2}
\end{equation}
where  $\chi^{(-)}_{cT}$ is the distorted wave generated by the potential
$U_{c}$.
The above matrix
element is dominated by small $x-c$ separations, where $\Psi^{CDCC}$
is at its best.

Although very appealing from the formal point of view, expression (\ref{Tpost2})
is hard to implement in practice. The main reason is that
this expression
involves a six-dimensional integral, in which   
both the initial and final wavefunctions  are highly
oscillatory. Furthermore, post form representations offer poor convergence
since both the scattering waves for $x+c$ and the potential $V_{xc}$ are
expressed in the same coordinate and consequently there maybe no
natural cutoff for the integral (\ref{Tpost2}) \cite{Ich85}. 

In order to make the calculation more feasible, Shyam and 
collaborators (see, for 
instance, \cite{Ban98,Chat00,Chat02})  have developed an approach
based in the amplitude (\ref{Tpost2}), in which the exact  wavefunction
is replaced by its elastic component, i.e.,
\begin{equation}
\Psi_{i}^{(+)}\approx \chi_{0}^{(+)}(\mathbf{R}_{AT})\phi_{0}(\mathbf{r}_{xc}),
\label{shyam}
\end{equation} 
where $\phi_a(\mathbf{r}_{xc})$ is the projectile ground state and 
$\chi_{0}^{(+)}$ is a distorted wave generated with a optical potential,
typically adjusted to  reproduce the elastic scattering data. Note that
this approximation neglects breakup events that proceed via projectile excitation.
Note also that, even after 
the replacement of the exact wavefunction by the elastic component,
Eq.~(\ref{Tpost2}) still involves a six dimensional integral. A significant
simplification of the problem can be achieved by using the local
momentum approximation \cite{Shy85,Chat00}, which leads to a factorization of the
amplitude in a product of two terms, each involving a three-dimensional integral.

The difficulties outlined above can 
be partially avoided using the prior representation of the transition 
amplitude,
\begin{equation} 
T_{prior}=
\langle \Psi _{f}^{(-)}|
V_{xT}+U_{c}-U_{AT}|\phi_{0} \chi_{0}\rangle ,
\label{T-prior}
\end{equation}
where the final state is the exact three-body wavefunction with
incoming boundary conditions. This wavefunction
is typically expanded in terms of the $x+T$ continuum states,
formally similar to Eq.~(\ref{cdccwf}), but now in the Jacobi set (2) of Fig.~\ref{jacobi}:
\begin{eqnarray}
\Psi_f^{(2)}({\bf r}_2,{\bf R}_2) = \sum_{\beta=0}^{N_2}
\phi_{\beta}({\bf r}_2) \chi_{\beta}({\bf R}_2).
\label{cdccwf2}
\end{eqnarray}
Again, the functions  $\phi_{\beta}$ represent the set of
bin wavefunctions \cite{Yah82,Yah86,Aus87}, constructed by superposition of pure scattering waves.
Note that Eq.~(\ref{cdccwf2}) goes beyond the DWBA, because couplings between final 
states are explicitly considered in the wavefunction $\Psi_f^{(2)}$.

Therefore, in the standard CDCC method the three-body 
continuum  is described in terms of the projectile two-body ($x-c$) 
states, while in the
amplitude (\ref{T-prior}) this continuum is expanded using the fragment-target 
states $x-T$.  
While the CDCC method treats the breakup process as inelastic excitations to the
projectile continuum, expressions (\ref{Tpost2}) and (\ref{T-prior}) emphasize a rather different
picture, in which three-body breakup is formally treated as transfer of one
of the fragments ($x$ in our case) to the target continuum. This is 
schematically illustrated in Fig.~\ref{coupbu}(b). 
At this stage, it is worth to stress that,
in the way here presented, the CDCC and the the TR* methods are solutions of the same three
three-body model Hamiltonian, given by Eq.~(\ref{H-cdcc}) and, therefore, three-body observables 
obtained with these two approaches should be the same.

However, in practice there are
several factors that may destroy this equivalence. First, due to computational
limitations, one can not include an arbitrarily large number of 
continuum states. Secondly, there are ambiguities associated
with the choice of the interactions involved in both schemes. In
the BU approach, one usually has two complex potentials, namely $U_x$ and
$U_c$, and a real interaction, $V_{xc}$. By contrast, in the amplitude
(\ref{T-prior}) the wavefunction $\Psi_f^{(2)}$ is typically obtained with
the complex potentials $U_c$ and $U_{xc}$. The choice of the potential $V_{xT}$
deserves special care. For inclusive processes, in which the fragment $x$ is
allowed to interact in any possible way with the target, $V_{xT}$ would
be a complicated many body operator, which can induce excitations in both 
$x$ and $T$. However, for a comparison with CDCC, in which only the elastic
breakup component is calculated, this operator is better represented by an
effective complex optical potential \cite{Ich85,Aus87} and hence, according
to our previous notation, $V_{xT}=U_x$. This is actually
the choice made in current semiclassical applications of the transfer to the continuum method
\cite{Bon01}. 



Another ambiguity is related to the  interaction that should be used for $U_{AT}$ in 
the amplitude (\ref{T-prior}). Note that, if the exact expression is used for the
three-body wavefunction $\Psi_f^{(-)}$ the matrix element is independent of the choice of the 
potential $U_{AT}$. This result does not hold when $\Psi_f^{(-)}$ is replaced by an 
approximated wavefunction. Following 
the standard DWBA choice, one could use the optical potential that reproduces the elastic 
scattering. Another possible choice, is the so called cluster-folding potential, given by
the sum of the fragments-target interactions folded with
the ground state of the projectile:
$\langle \phi_0|
U_{x}  + U_{c}|\phi_0\rangle$. In our calculations we have explored both choices.

The main purpose of this work is to test to what 
extent the equivalence between the BU and (prior form) TR* is 
satisfied, at least in an approximate way,
in actual calculations. To this end, we have performed numerical calculations
for two different systems using both approaches, and compared several reaction observables. At 
high scattering energies, around 100 MeV per nucleon and above, the TR* method, as 
presented here, becomes
numerically very demanding, and the problem is better solved by using further approximations, such
as the use of classical trajectories. Since it is our purpose to compare the full quantum mechanical
CDCC and TR* expressions, we confine ourselves to reactions at low and medium  energies.

\section{Calculations}
\subsection{$p$+$^{11}$Be case}
We first consider the reaction of 38.5 MeV per nucleon $^{11}$Be breaking up on protons.
The elastic and transfer channels were measured in GANIL 
\cite{Win01,lapoux} but no breakup data was recorded. 
According to the discussion in the previous section, the 
$^{11}$Be breakup reaction can be thought of as
the direct breakup (BU) $^{11}$Be$+p \rightarrow (^{10}$Be$+n) +p$ or
transfer of the neutron to the continuum of the deuteron (TR*)
$^{11}$Be$+p \rightarrow ^{10}$Be$+ (n+p)$. 

The $n-p$ interaction was taken from \cite{Aus87}, whereas the nuclear 
interaction
for $p-^{10}$Be was extracted from a fit to the elastic data \cite{lapoux}. 
The Coulomb potential for  $p-^{10}$Be was also included so Coulomb breakup
is also included in our calculations, although it was shown to be very small.
The binding potential
and the potential generating the continuum waves for $n-^{10}$Be 
was the same as in \cite{Win01}, but without the spin-orbit term. These
potentials are listed in Table \ref{Tab:OP}. The BU 
calculations required partial waves up to 
$l_{max}=4$ and energies up to $\varepsilon_{max}$=30 MeV for the relative 
motion of the $n-^{10}$Be system. The bin wavefunctions for the CDCC couplings
were calculated up to $R_{bin}$=60 fm. 
An $L_{max}=25$ was necessary for the $^{11}$Be$-p$ distorted waves.  
As to the TR* calculation, the same parameters were sufficient for convergence
but they are computationally more lengthy
\footnote{The converged BU calculation required 48 Mb RAM and 
approximately 10 minutes in a linux 2.4 Ghz PC and
the TR* calculation used 1.4 Gb RAM and took approximately 30 min on the
same machine.}.
All the TR* calculations here presented use as incoming optical potential the 
folding of the $p-n$ and $p$-$^{10}$Be interactions with the ground state wavefunction
of the  $^{11}$Be nucleus. We also did calculations using a Woods-Saxon shape
with the same parameters as for the $p$-$^{10}$Be potential. Results obtained 
with this potential
are very similar to those of the cluster-folding and, hence, will 
not be shown in the graphs. Both the BU and TR* calculations were performed with the
computer code FRESCO \cite{Thom88}.

\begin{table}
\caption{\label{Tab:OP}Optical model parameters used in this work. Except
for the $p+n$ case, all potentials are parameterized using the usual Woods-Saxon
form, with a real volume part and volume ($W_v$)  or surface ($W_d$) 
imaginary part. Reduced
radii are related to physical radii by $R=r_0 A_T^{1/3}$.}
\begin{ruledtabular}
\begin{tabular}{c|ccccccc}
System &
\( V_{0} \)&
\( r_{0} \)&
\( a_{0} \)&
\( W_{v} \)&
\( W_{d} \)&
\( r_{i} \)&
\( a_{i} \)
\\
 &
(MeV) &
(fm)  &
(fm)  &
(MeV)  &
(MeV)  &
(fm)  &
(fm) \\
\hline
$p$+$^{10}$Be&
51.2&
1.114&
0.57&
19.5&
0&
1.114&
0.50 
\\
$p+n$\footnote{Gaussian geometry: $V(r)=V_0 \exp[(r/r_0)^2]$.}&
72.15&
1.484&
-&
-&
-&
-&
- 
\\
 & & & & & & \\
$^{8}$B+$^{58}$Ni&
130 &
1.050&
0.65&
92&
0 &
1.123&
0.997\\ 
$^{7}$Be+$^{58}$Ni &
100 &
1.050 &
0.65 &
30.6 &
0    &
1.123 &
0.80 \\
$p$+$^{58}$Ni &
54.512 &
1.17 &
0.75 &
0    &
11.836 &
1.260    &
0.58  \\
$p$+$^{7}$Be \footnote{In the TR* case, this potential 
includes also an imaginary part with the 
same geometry as the real part.} &
44.675 &
1.25 &
0.52 &
- &
- &
- &
- \\
\end{tabular} 
\end{ruledtabular}
\end{table}

\begin{figure}
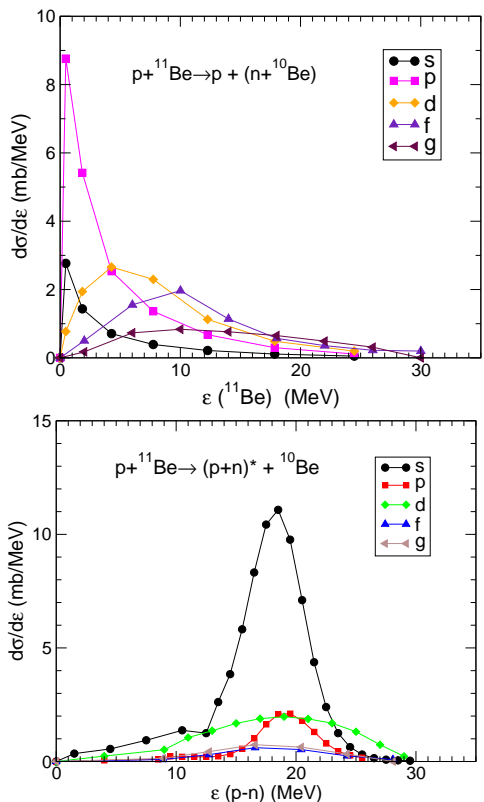

\resizebox*{0.35\textwidth}{!}{\includegraphics{be11pd_bu.eps}}
\resizebox*{0.35\textwidth}{!}{\includegraphics{be11pd_tc.eps}}
\vspace{-0.4cm}
\caption{\label{be11_vsE} (Color online) Breakup energy distribution for   
$^{11}$Be on protons at 38.5 MeV/u:
comparison of transfer to the continuum (bottom figure) 
with the direct breakup approach (upper figure).}
\end{figure}

In Fig.~\ref{be11_vsE} we present the differential breakup cross 
section calculated within the BU and TR* methods, as a function
of the excitation energy of the  $^{10}$Be-$n$ and  $p$-$n$ systems,
respectively. It can be seen that, in the BU case,  
most of the strength is below $\varepsilon_x(^{11}$Be$)\approx 5$ MeV whereas
in the TR* calculation the strength is largely
concentrated around $\varepsilon_x(d) \approx 20$ MeV. 
The total integrated cross  section for the two processes
are $\sigma_{bu}=125$ mb and $\sigma_{tr}=140$ mb.

\begin{figure}
\resizebox*{0.4\textwidth}{!}{\includegraphics{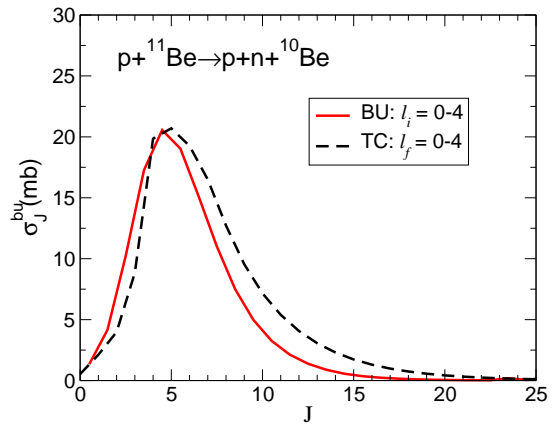}}
\vspace{-0.4cm}
\caption{\label{be10J} Total angular momentum distribution for the  
breakup of $^{11}$Be on protons at 38.5 MeV/u:
comparison of transfer to the continuum with the direct breakup 
approach.}
\end{figure}
In order to establish a meaningful comparison between the two approaches,
one has to compare the same quantities. For this purpose,
in Fig.~\ref{be10J} we compare the contribution 
of each total angular momentum $J$ 
(resulting from the vector coupling of projectile, target and their relative
motion angular momentum) to the breakup cross section.
It can be seen that both 
distributions are  similar for small values of $J$.
Also, we find that for the two cases the distribution
peaks around $J=5$ which means that most of the breakup cross section
occurs at  distances $b \approx 4$~fm. The similitude between both
distributions supports the idea
that breakup, calculated as excitation of the projectile to its continuum
spectrum, or by transfer to the continuum states of the target, do describe
the same physical process.
However, for $J>5$ the TR* clearly exceeds the BU cross section 
which, as we will show below, results on different predictions for measurable 
physical observables.

\begin{figure}
\resizebox*{0.4\textwidth}{!}{\includegraphics{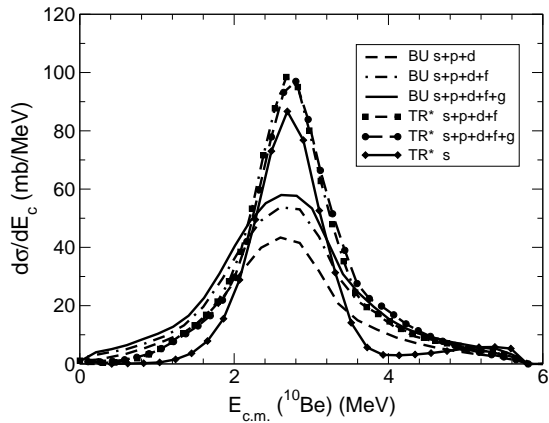}}
\vspace{-0.4cm}
\caption{\label{be10e} Energy distribution in the overall c.m. for the $^{10}$Be
coming from the breakup of $^{11}$Be on protons at 38.5 MeV/u:
comparison of transfer to the continuum with the direct breakup 
approach.}
\end{figure}

In actual breakup experiments, the data commonly recorded are the angular and/or
energy distributions of the emerging fragments. Therefore, it is instructive
to compare the predictions of both approaches for these  observables.
In Fig.~\ref{be10e} 
we represent the calculated breakup cross section distribution 
of the outgoing $^{10}$Be fragments as a function of its kinetic energy, measured
in the overall c.m.\  of the three-body system. In both methods, these distributions
are obtained by integration of a triple differential cross section with respect to the 
angular variables.
In the case of the TR* approach, this procedure is straightforward, since  
expression (\ref{T-prior}) is referred already to the scattering angle and energy of the
$^{10}$Be fragments. 
In the case of the BU approach, the  differential cross section is naturally
expressed in terms of the scattering angle of the composite $x+c$ and the relative
energy between these two fragments. In order to obtain the differential cross section
with respect to any of the fragments, one has to apply to appropriate kinematic 
transformation, as done in Ref.~\cite{Tos01}.

These energy distributions show only a qualitative agreement between the two
methods. In both cases, the energy of the $^{10}$Be fragments goes from
zero to about 6 MeV, with a maximum around 3 MeV. However, although 
the same energy region
of space is being included, the two models do not produce
identical shapes. In Fig.~\ref{be10e} we also show the convergence rates
for both TR* and BU.  The labels s,p,d, etc refer to the relative partial
waves included in the corresponding calculation. For instance,
the solid line in the BU calculation includes all $n-^{10}$Be partial waves
up to $l=4$ whereas the dashed line includes only partial waves up to $l=2$.

\begin{figure}
\resizebox*{0.4\textwidth}{!}{\includegraphics{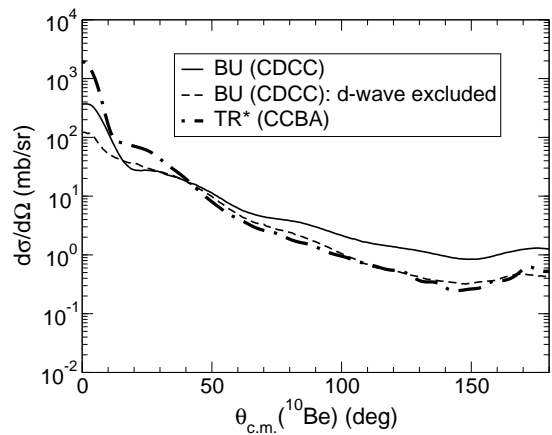}}
\vspace{-0.4cm}
\caption{\label{be10ang} Angular distribution in the c.m.\ for the $^{10}$Be
coming from the breakup of $^{11}$Be on protons at 38.5 MeV/u:
comparison of transfer to the continuum with the direct breakup 
approach.}
\end{figure}

Disagreements become more severe  for the
angular distributions. These can be seen in Fig.~\ref{be10ang}.
Note that the TR* calculation (dot-dashed curve in Fig.~\ref{be10ang})
exhibits a pronounced decrease of the cross section as a function of angle. 
In addition, its forward angle cross section is an order of magnitude larger 
than the BU calculation (solid line) and an order 
of magnitude lower for backward angles.
Part of the reason for the disagreement can be understood excluding the
d-wave resonance in the $n-^{10}$Be system 
(dashed curve). 
This wave has a very strong
contribution for backward angles and a d-wave resonance in $^{11}$Be
will be very hard to model in terms of the deuteron continuum. 
However, the discrepancy remains at forward angles: including only the
s-wave of the deuteron in the BU calculation, the resulting
cross section is an order of magnitude
smaller than the TR* cross section.
Detailed data for this reaction would be very useful.


\subsection{$^8$B+$^{58}$Ni case}
A breakup reaction for which more detailed data exist is that for
$^8$B$\rightarrow ^7$Be$+p$ on $^{58}$Ni at 25.6 MeV. 
Calculations using the standard CDCC 
$^{8}$B$+^{58}$Ni $\rightarrow (^{7}$Be$+p) +^{58}$Ni
have provided very good agreement with experiment \cite{Nun99,Tos01}.
Again, one can think of the alternative path to breakup, as 
transfer to the continuum of the $^{59}$Cu nucleus (TR*)
$^{8}$B$+^{58}$Ni $\rightarrow ^{7}$Be$ +(p+^{58}$Ni). 
All interactions for both BU  and TR* are the same as those in \cite{Tos01},
although for the $p-$Ni only the real part was included in the 
TR* calculation.

The BU calculations required partial waves up to 
$l_{max}=4$ and energies up to $\varepsilon_{max}=8$ MeV for the relative 
motion of the $p$-$^{7}$Be system. The bin wavefunctions for the CDCC couplings
were calculated up to $R_{bin}=60$ fm and the coupled channel equations
were solved with $R_{max}=500$ fm. An $L_{max}=1000$ was necessary for the 
$^8$B$-^{58}$Ni distorted waves
\footnote{The converged TR* calculation required
400 Mb RAM and took approximately 8 hours.}.
Note that in this case the BU calculation is a good test 
reference, as it agrees
very well with both energy and angular distribution data \cite{Tos01}, at
least within the kinematic conditions of the referred experiment.
As to the TR* calculation we used  partial waves up to
$l_{max}=17$ and energies up to $\varepsilon_{max}=10$ MeV for the 
relative motion $p+^{58}$Ni. To reduce the computational requirements,
for $l_{f}> 6$, continuum--continuum  couplings 
were included only between bins with the same $l_f$. 
The bin wavefunctions 
were calculated up to $R_{bin}=120$ fm, and 
an $L_{max}=120$ was necessary for the 
$^8$B$-^{58}$Ni distorted waves.  However, these  results are not yet converged.
The large required widths for the non-local transfer couplings
make the calculations extremely heavy. 

We next compare the same 
quantities as in the $p$+$^{11}$Be case. Unlike the previous test example,
where $V_{xc}=V_{pn}$, here
the $V_{xc}$ interaction ($p$+$^{7}$Be), as extracted from the elastic data, 
is expected
to contain an imaginary part. In our TR* calculations, we probe several possibilities for
the imaginary part, keeping the same geometry as the real part and using different
choices for the depth.
For the incoming
channel optical potential we used two different potentials. The first one, denoted
OM1, is the sum of the $p$+$^{7}$Be and $^{7}$Be+$^{58}$Ni interactions folded with the 
bound state wavefunction of the $^{8}$B nucleus. 
The second one, consisted on a parametrization with 
two Woods-Saxon terms, real and imaginary, with parameters obtained by fitting
the elastic angular distribution, as predicted by the CDCC calculation. In this case we
used, as starting parameters for the fitting routine, those for the $^{7}$Be+$^{58}$Ni 
interaction. The parameters for this potential resulting
from the fit, denoted OM2, are listed in Table \ref{Tab:OP}.

\begin{figure}
\resizebox*{0.4\textwidth}{!}{\includegraphics{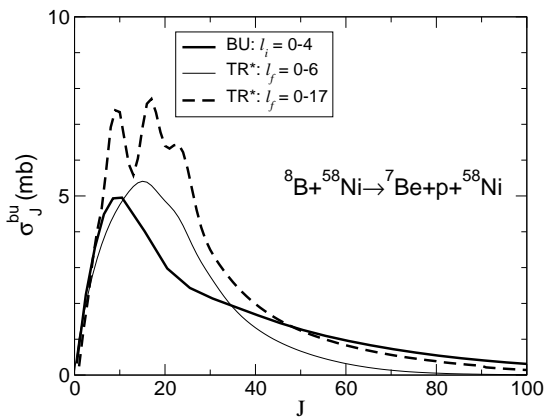}}
\vspace{-0.4cm}
\caption{\label{b8J} Total angular momentum distribution for the  
breakup of $^{8}$Be on $^{58}$Ni at 25.6 MeV. The solid line
is the BU calculation, whereas the remaining lines 
correspond to the TR* calculations with two different sets of partial waves $l_f$,
as indicated by the labels (see text for details).}
\end{figure}

In Fig.~\ref{b8J} we show the total breakup calculated within the BU
and TR* schemes, for each value of the total angular momentum, $J$. In 
the TR* case,  two different calculations are presented. In both cases,
the potential OM2 was used for the elastic channel, and the value $W_d$=3 was
used for the imaginary depth of the $p$+$^{7}$Be interaction.
The thin solid line in this figure represents the TR* calculation performed in the subspace
$l_f=0-6$. This calculation exhibits clear differences from the BU (thick solid line): the 
lower values of $J$ are clearly overestimated, whereas for the large values of $J$,
 the distribution
falls too fast as compared to the BU. 
The second  TR* calculation here presented (dotted-dashed line)  uses the same parameters
as before, but includes also the partial waves $l_f=7-17$ for the proton-$^{58}$Ni
continuum. This calculation improves the agreement for large $J$. However, it also
adds an extra contribution on the lower values of $J$ which appears to deteriorate the
agreement with the BU results.

As noted in the previous section,  it has been argued that in the calculation
of elastic breakup, one should use an absorptive potential for
the $V_{xT}$ operator in Eq.~(\ref{T-prior}). This might explain part 
of the discrepancies found here between both  approaches. Unfortunately, the present 
version of the code {\tt FRESCO} does not allow
this interaction to be complex.
Although at present we cannot use an imaginary term in the $V_{xT}$ operator,
we are considering modifications of the code to enable this feature.
It is fortunate that, in
the $p$+$^{11}$Be reaction, $V_{xT}=V_{pn}$ which, within the energy range of our analysis,
is well represented by a real potential. This might explain the better agreement obtained
for the absolute value of the breakup cross section, as well a for the $J$-distribution, as 
compared to the $^{8}$B+$^{58}$Ni case.

Although the representation of the $V_{xT}$ operator by a real operator
has undesirable consequences for the purpose of the present work,
one may speculate about the physical meaning behind this choice.
As noted above, choosing $V_{xT}$ as the phenomenological optical
potential in the CDCC approach implies that the model space considers
only breakup where the target is left in its ground state.
Conversely, one may argue that, by choosing this potential as real,
as we do here in the TR* case,  there is the  possibility of
including other inelastic breakup events, and even transfer to
bound states of the target. In other words,
the TR* method with a real $V_{xT}$ may include
contributions from the so called stripping breakup.
These aspects will be explored in future work.

\begin{figure}[t]
\resizebox*{0.4\textwidth}{!}{\includegraphics{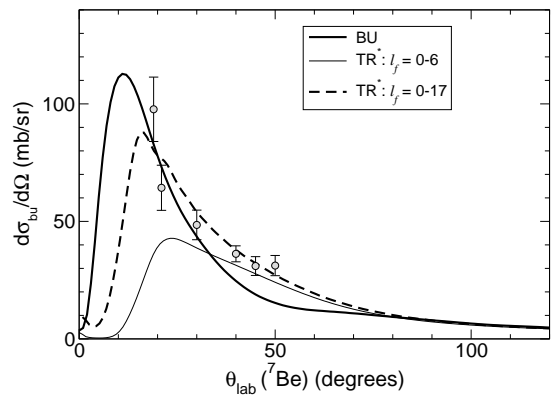}}
\vspace{-0.4cm}
\caption{\label{be7ang_lf} (Color online) Angular distribution in the laboratory frame 
for the $^{7}$Be
fragments coming from the breakup of $^{8}$B on $^{58}$Ni at 25.6 MeV
within the direct breakup and transfer to the continuum approaches. Experimental data
are from  \cite{Gui00}.}
\end{figure}
%
For the  $^{7}$Be angular distribution, we first study the convergence of the
calculations with respect to the size of the model space. In the case of the
BU  this has been discussed in detail in Ref.~\cite{Tos01}, and so we will  
just quote the results from this reference. In this section,
we concentrate only on the convergence of this observable within the TR* 
scheme. For  definiteness, these calculations were performed with 
the choice $W_d=3$ MeV for the imaginary part of the  $p$+$^{7}$Be potential. The 
dependence on this potential will be analyzed below.
The convergence of the calculation with respect to the number of partial
waves  for the relative motion of the $x-T$ pair is depicted in 
Fig.~\ref{be7ang_lf}. For comparison, the BU calculation (thick solid line) and the
experimental data points \cite{Gui00} have been 
also included. The thin solid line is  the calculation where the 
partial waves $l_f=0-6$ are included and coupled among them to all orders. The thick
dotted-dashed line is the sum of this calculation and the separated differential 
cross sections for the
partial waves  $l_f=7-17$. It becomes apparent that the contribution of these
partial waves is very important to describe the strong Coulomb peak at 
small scattering angles. Both calculations are in good agreement with the data. 
Given the large error bars and restricted angular range
of the data it is not possible to make strong conclusions on which method is more 
suitable in this particular situation. Roughly speaking, it seems that the TR* 
describes better the larger angles, while the BU is more suitable to describe the
smaller angles.
In the remaining discussion, all our comparisons 
with the BU calculations
will be performed with the full set of partial waves ($l_f=0-17$).

\begin{figure}[t]
\resizebox*{0.4\textwidth}{!}{\includegraphics{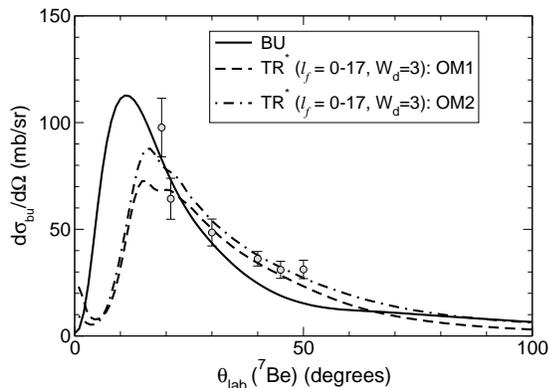}}
\vspace{-0.4cm}
\caption{\label{be7ang_om1} Angular distribution in the c.m. for the $^{7}$Be
fragment coming from the breakup of $^{8}$B on $^{58}$Ni at 25.6 MeV:
comparison of transfer to the continuum with the direct breakup 
approach for two different choices of the incoming optical potential for
the TR* calculation (see text for details).}
\end{figure}
Next, we study the dependence of the $^{7}$Be angular distribution 
on the choice of the incoming channel
optical potential. This is illustrated in Fig.~\ref{be7ang_om1}.
 In this figure, the dashed and 
dotted-dashed lines correspond, respectively, to the TR* calculation with 
the cluster-folded potential (OM1) and
the phenomenological optical potential obtained from a fit of the CDCC 
elastic angular distribution (OM2).
Again, we have  fixed 
the imaginary depth of the $p$+$^{7}$Be interaction to $W_d=3$~MeV. 
For comparison purposes, the
BU calculation (thick solid line) has been also included. One sees that the choice of 
the elastic channel optical potential has indeed an effect on the predicted $^{7}$Be
cross sections. However, given the uncertainties of these calculations, differences do not
seem very dramatic, and in both cases a fairly good agreement is obtained with
the BU calculation irrespective of the choice of this potential. 

\begin{figure}[t]
\resizebox*{0.4\textwidth}{!}{\includegraphics{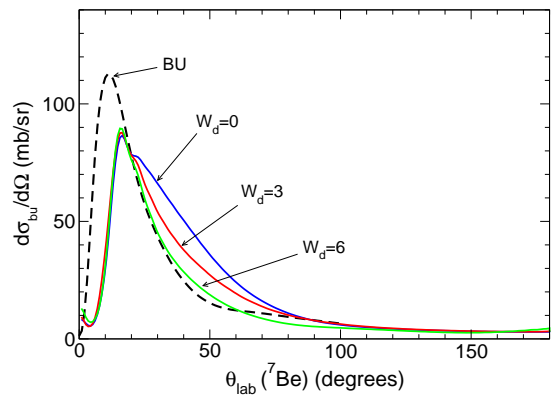}}
\vspace{-0.4cm}
\caption{\label{be7ang_Up} (Color online) Angular distribution 
for the $^{7}$Be
fragments coming from the breakup of $^{8}$B on $^{58}$Ni at 25.6 MeV:
comparison of transfer to the continuum with the direct breakup 
approach for different values of the imaginary depth of the $p$+$^{7}$Be potential.}
\end{figure}

In Fig.~\ref{be7ang_Up} we compare the 
standard CDCC calculation with  TR* calculations performed with the optical potential
OM2 for the incoming channel, and different choices of the
imaginary depth  of the $p$+$^{7}$Be potential. Thick lines correspond to the full calculation
($l_{max}=17$). 
At backward angles all TR*
calculations look  very similar, indicating a fast convergence with respect to the number
of partial waves and a weak dependence on the choice of the $p$+$^{7}$Be potential at these
angles. The effect of the imaginary part seems to be crucial at intermediate angles, where
one observes a progressive suppression of the cross section with increasing absorption.
The TR* with real $p$+$^{7}$Be interaction clearly overestimates the BU result at intermediate
angles. Interestingly, the forward angular region is only weakly affected by 
this absorptive term.
As we verified in our calculations, this is a consequence of the fact that 
this potential has little effect on the higher partial waves.
The best agreement with the BU calculation is obtained when the imaginary part 
of the $p$+$^{7}$Be potential for TR* is $W_d=6$ MeV.


The reasonably
good agreement between BU calculation  and the TR* calculations, performed in the augmented model 
space ($l_{max}=17$) and with a complex proton+$^{7}$Be interaction, leads us again to the
conclusion that projectile breakup  and transfer to the target continuum 
populate, to a large extent, the same three-body continuum. 
We interpret the discrepancy at small scattering angles as lack of convergence
of our TR* calculations, and the ambiguities associated to the potentials.

\begin{figure}[b]
\resizebox*{0.4\textwidth}{!}{\includegraphics{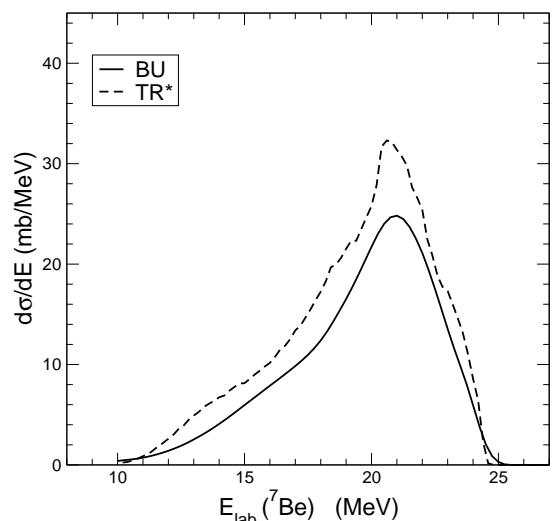}}
\vspace{-0.4cm}
\caption{\label{be7e} Energy distribution in the c.m. for the $^{7}$Be
coming from the breakup of $^{8}$B on  $^{58}$Ni at 25.6 MeV:
comparison of transfer to the continuum with the direct breakup 
approach.}
\end{figure}

In Fig.~\ref{be7e}, we plot the energy distribution for the
detected $^7$Be fragments after breakup, for the BU (solid line) and the TR* approach with
$l_{max}=17$, and $W_d=3$ MeV for $p$+$^{7}$Be (dashed line). 
It can be seen that both methods give  similar distributions. In particular, it is
noticeable that both methods predict a maximum of the energy distribution at about the same
$^{7}$Be energy. The TR* calculation gives however a larger breakup cross section.

\begin{figure}
\resizebox*{0.4\textwidth}{!}{\includegraphics{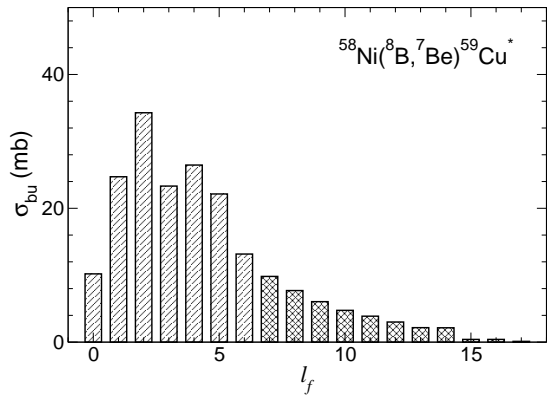}}
\vspace{-0.4cm}
\caption{\label{be7lf} Breakup distribution as a function of  
the relative motion $^{58}$Ni-p
for the transfer to the continuum of the proton from $^{8}$B 
to $^{58}$Ni at 25.6 MeV. Results up to $l_f=6$ are fully coupled,
but for $l_f>6$ they are calculated separately
(see text for details).}
\end{figure}

To have further insight into the convergence of the TR* calculation with
respect to the size of the model space  we have plotted in Fig.~\ref{be7lf}
the distribution of the TR* cross section as a function of  $l_f$, i.e., the
final angular momentum between the proton and the target.
On one hand, it is clear that TR* requires far more partial waves that the
BU calculation. On the other hand, the small
contribution for $l_f>15$, does provide some confidence in the results
presented here.  
The breakup of $^8$B on $^{58}$Ni at 25.8 MeV 
is a good example where the BU configuration seems to work
better than the  TR* configuration.

\section{Discussion}
The qualitative agreement between the calculations performed in the BU and TC representations
clearly indicates that both basis describe to a large extent the same three-body
continuum. However, the analysis of the preceding section also shows that, in order
to achieve convergence of the observables, the number of basis states required in 
both representations can be very different. From the practical point of view, it will be
desirable in general to choose the representation that requires less number of states.

For example, our analysis of the  $^8$B breakup reaction clearly supports 
the choice of the Jacobi coordinates (1) in Fig~\ref{jacobi}. On the other side, 
our previous study on 
the reaction $^8$Li+$^{208}$Pb \cite{Mor03a} was better performed using 
the coordinate set (2). It is clear that, in general, 
the most suitable choice will depend on the specific reaction. 
Inspired by the work of Merkuriev \cite{Merk74} on three-body bound states, 
we have searched for a criterion that can select between the two 
representations. Unfortunately the asymptotic behaviour of the three 
body continuum is very different from the exponential decay of bound states,
and the final behaviour of these expansions is not as transparent.
Therefore, we will present only qualitative arguments 
to evaluate the relative 
importance of the different configurations.

For practical purposes one always opts for the calculation that requires
the minimum number of partial waves in the $x-c$ or $x-T$ systems, for 
BU and TR* respectively. For the $^{11}$Be 
the difference in $l_{max}$ for BU and TR*  was not noticeable.
For the $^8$B example, $l_{max}=4$ was sufficient for BU whereas
$l_{max}=17$ was still not enough for the TR*.
In addition to the angular momentum, it seems clear that if a 
representation such as Eq.~(\ref{cdccwf})
is valid, then the average energy $\langle\varepsilon_1\rangle$ 
associated with the relative
coordinate $r_1$ should be
much less than the total energy in the centre of mass frame $E_{cm}^{(1)}$. 
Equally, if Eq.~(\ref{cdccwf2}) is to be used, the average energy 
$\langle\varepsilon_2\rangle$ associated with 
coordinate $r_2$ should be small compared 
to the total energy of the exit partition in the centre of mass frame 
$E_{cm}^{(2)}$. We have computed the average
energy between the fragments in the continuum, weighting it
by the cross section. For the $^{8}$B example
above, we obtain $\langle\varepsilon_1\rangle=1.85$ MeV with $E_{cm}^{(1)}=22.7$ 
MeV for the BU calculation and $\langle\varepsilon_2\rangle=7.84$ MeV with 
$E_{cm}^{(2)}=26.0$ MeV 
for the TR* calculation.
In the first case $\langle\varepsilon_1\rangle/E^{(1)}_{cm}=0.08$ whereas 
in the latter
$\langle\varepsilon_2\rangle/E^{(2)}_{cm}=0.30$. As to the $^{11}$Be example, 
the difference
is also pronounced: for BU we obtain 
$\langle\varepsilon_1\rangle/E^{(1)}_{cm}=0.15$ and 
$\langle\varepsilon_2\rangle/E^{(2)}_{cm}=0.52$,
implying again that the transfer to the continuum approach is
not the best. In the breakup of $^8$Li \cite{Mor03a}, the TR* approach was
used successfully. We compute the average energy  between
the fragments in the outgoing channel and obtain 
$\langle\varepsilon_2\rangle/E^{(2)}_{cm}=0.05$,
validating the previous TR* calculations \cite{Mor03a}.  
This same criterion shows a red card to the preliminary calculations
on $^6$He \cite{Agu00}, as in that case we have 
$\langle\varepsilon_2\rangle/E^{(2)}_{cm}=0.43$.
In conclusion, the  condition $\langle\varepsilon^i\rangle/E^i_{cm}\ll1$ 
should be satisfied
whenever only one $i$ Jacobi set is taken into account in the 
reaction formalism.
This is for instance the case in 
Coulomb dissociation, where the long-range and smooth behaviour of 
the Coulomb potential makes that the reaction mechanism populates mainly low-energy 
states of the projectile. 
As a matter of fact, is was shown in \cite{Nun99} that the peak
in the breakup angular distribution at $\theta_\mathrm{c.m.}\approx 15^{\circ}$ is mainly
due to  Coulomb excitation. In our calculations, this peak is well reproduced by the CDCC
calculation, while it requires many partial waves of the proton-target system in the TR* 
method. Furthermore, at high energies, the Coulomb dissociation cross section becomes 
approximately proportional to the $B(E1)$ strength of the projectile. Hence, the 
BU approach provides in this case a more transparent and useful picture of the reaction
process. 
On the other side, in situations where the removed particle has a high probability
to be left with a small relative energy with respect to the target, the TR* may result more
convenient. Nevertheless, this does not seem to be the case of the reactions studied in 
this work.

Even in calculations where the TR* converges quickly with the number of partial
waves, we found it to be computationally very demanding, as it 
involves non-local couplings. In practice, these calculations
could be significantly speeded up, using different techniques which are now of common
use: local momentum and adiabatic approximations, etc. 
It is  however beyond the scope of this work to explore how to make the 
TR* numerically more feasible.

\section{Conclusions}
Given the importance of the reaction model in the understanding
of the fundamental nuclear structure on the drip lines, we 
compare two alternative schemes to calculate breakup observables for
the reaction $A(=c+v)+T$, within the same three-body Hamiltonian. Each one of these methods uses  
a description of the three-body continuum in terms of one 
of the possible sets of Jacobi coordinates. In the 
CDCC approach, the three-body continuum is described in terms of the $c-v$ states. On the
contrary, the transfer to the continuum (TR*) approach expands the continuum in 
terms of $v-T$ states. Since both sets of states form a complete basis, reaction
observables could be in principle calculated using either of these two basis.
We show that in both cases predictions by these two schemes are in semiquantitative agreement. 
This result clearly shows that, provided that
enough basis states are included, both representations describe essentially the same three-body
continuum. 
In the $^8$B+$^{58}$Ni 
reaction, both calculations are consistent with existing experimental data. 
From our analysis, it is clear that the truncated model space
is not always identical. 
In particular, in this case we found that the TR* approach 
requires a significantly larger number of partial waves for the proton-cluster relative
motion, thus making the 
calculation numerically more demanding.
We also find that part 
of the disagreement between the two methods is due to the ambiguities associated with the choice of
the effective interactions involved in both methods.
In addition, we have proposed a simple criterion based on the average relative excitation energy 
to select between the two approaches.

More detailed studies 
on the absorption part of the optical potential and applications
to other reactions will be presented elsewhere. Ultimately, we
would like to compare these results with exact Faddeev calculations.
Work along these line is being initiated.

\begin{acknowledgments}
This work has been partially
supported by Funda\c{c}\~ao para a Ci\^encia e a Tecnologia (F.C.T.)
of Portugal, under the grant POCTIC/36282/99 and by the NSCL at Michigan State
University. One of the authors (A.M.M.) acknowledges a 
research grant by the Junta de Andaluc\'{\i}a (Spain).
We are deeply grateful to Prof.\ J.\ Tostevin for providing us with his code
to calculate 3-body  observables and for his assistance in using it.
\end{acknowledgments}

\bibliographystyle{unsrt}
\bibliography{./refer}
\end{document}